\documentclass[12pt]{article}

\usepackage[utf8]{inputenc}
\usepackage{textcomp}
\usepackage{chetnew}
\usepackage{tikz}
\usepackage{amssymb,amsfonts,mathrsfs,dsfont,yfonts,bbm}\usetikzlibrary{calc}
\usepackage{xcolor}
\usepackage{braket}

\newcommand{\CC}{\mathcal{C}}
\newcommand{\CO}{\mathcal{O}}
\newcommand{\CT}{\mathcal{T}}
\newcommand{\CI}{\mathcal{I}}
\newcommand{\CN}{\mathcal{N}}

\newcommand*{\boxcoloro}{orange}
\makeatletter
\newcommand{\boxedo}[1]{\textcolor{\boxcoloro}{%
\tikz[baseline={([yshift=-1ex]current bounding box.center)}] \node [rectangle, minimum width=1ex,rounded corners,draw] {\normalcolor\m@th$\displaystyle#1$};}}
 \makeatother
\newcommand*{\boxcolorr}{red}
\makeatletter
\newcommand{\boxedr}[1]{\textcolor{\boxcolorr}{%
\tikz[baseline={([yshift=-1ex]current bounding box.center)}] \node [rectangle, minimum width=1ex,rounded corners,draw] {\normalcolor\m@th$\displaystyle#1$};}}
 \makeatother
\newcommand*{\boxcolorb}{blue}
\makeatletter
\newcommand{\boxedb}[1]{\textcolor{\boxcolorb}{%
\tikz[baseline={([yshift=-1ex]current bounding box.center)}] \node [rectangle, minimum width=1ex,rounded corners,draw] {\normalcolor\m@th$\displaystyle#1$};}}
 \makeatother
\newcommand*{\boxcolorg}{green}
\makeatletter
\newcommand{\boxedg}[1]{\textcolor{\boxcolorg}{%
\tikz[baseline={([yshift=-1ex]current bounding box.center)}] \node [rectangle, minimum width=1ex,rounded corners,draw] {\normalcolor\m@th$\displaystyle#1$};}}
 \makeatother
 \newcommand*{\boxcolorp}{purple}
\makeatletter
\newcommand{\boxedp}[1]{\textcolor{\boxcolorp}{%
\tikz[baseline={([yshift=-1ex]current bounding box.center)}] \node [rectangle, minimum width=1ex,rounded corners,draw] {\normalcolor\m@th$\displaystyle#1$};}}
 \makeatother
  \newcommand*{\boxcolorc}{cyan}
\makeatletter
\newcommand{\boxedc}[1]{\textcolor{\boxcolorc}{%
\tikz[baseline={([yshift=-1ex]current bounding box.center)}] \node [rectangle, minimum width=1ex,rounded corners,draw] {\normalcolor\m@th$\displaystyle#1$};}}
 \makeatother
  \newcommand*{\boxcolory}{yellow}
\makeatletter
\newcommand{\boxedy}[1]{\textcolor{\boxcolory}{%
\tikz[baseline={([yshift=-1ex]current bounding box.center)}] \node [rectangle, minimum width=1ex,rounded corners,draw] {\normalcolor\m@th$\displaystyle#1$};}}
 \makeatother

\usepackage{amsmath}	
\begin{document}
\preprint{QMUL-PH-18-11}

\title{Flowing from 16 to 32 Supercharges}

\author{Matthew Buican$^{\diamondsuit,1}$, Zoltan Laczko$^{\clubsuit,1}$, and Takahiro Nishinaka$^{\heartsuit,2}$}

\affiliation{\smallskip $^{1}$CRST and School of Physics and Astronomy\\
Queen Mary University of London, London E1 4NS, UK\\ $^{2}$Department of Physical Sciences, College of Science and Engineering\\ Ritsumeikan University, Shiga 525-8577, Japan\emails{$^{\diamondsuit}$m.buican@qmul.ac.uk, $^{\clubsuit}$ z.b.laczko@qmul.ac.uk, $^{\heartsuit}$nishinak@fc.ritsumei.ac.jp}}

\abstract{We initiate a study of an infinite set of renormalization group flows with accidental supersymmetry enhancement. The ultraviolet fixed points are strongly interacting four-dimensional $\CN=2$ superconformal field theories (SCFTs) with no known Lagrangian descriptions, and the infrared fixed points are SCFTs with thirty-two (Poincar\'e plus special) supercharges.}

\date{July 2018}

\setcounter{tocdepth}{2}

\maketitle
\toc

\newsec{Introduction}
Emergent symmetries are ubiquitous in quantum field theory (QFT):\footnote{Throughout this note we use \lq\lq emergent" symmetries and \lq\lq accidental" symmetries interchangeably.} along renormalization group (RG) flows, couplings that break certain symmetries are sometimes renormalized to zero at long distance. The resulting infrared (IR) theory then has accidental symmetries that are not present in the ultraviolet (UV) theory.\footnote{In general, it is an interesting but difficult question to try to find constraints on the amount of accidental symmetry (e.g., see \cite{Abel:2011wv,Buican:2011ty,Buican:2012ec,Collins:2016icw} for a discussion in the context of certain classes of RG flows).}\footnote{Here we have in mind symmetries that act on local operators. One may generalize the concept of emergent symmetry to include higher-form symmetries as well (e.g., see \cite{Cordova:2018cvg}).}

Often, supersymmetry (SUSY) is one of these emergent symmetries. For example, in three dimensions, one may potentially find accidental $\CN=1$ SUSY in certain condensed matter systems \cite{Balents:2008vd,Grover:2013rc} (see also \cite{Lee:2006if} for a discussion of emergent $\CN=2$ SUSY).

More generally, additional SUSY can emerge in RG flows that are already supersymmetric. Instances of this phenomenon in three dimensions include the $\CN=3\to\CN=6$ (or $\CN=8$) enhancement in the ABJM flows starting from certain deformed super Yang-Mills (SYM) theories in the UV \cite{Aharony:2008ug} as well as $\CN=1\to\CN=2$ enhancement studied in other contexts \cite{Gaiotto:2018yjh,Benini:2018bhk} (see also \cite{Gang:2018huc} for a recent discussion in the context of 3D $\CN=2\to\CN=4$). In four dimensions, enhancement from $\CN=1\to\CN=2$ has also received considerable attention recently \cite{Gadde:2015xta,Maruyoshi:2016tqk,Maruyoshi:2016aim,Agarwal:2016pjo,Benvenuti:2017bpg,Giacomelli:2017ckh,Agarwal:2017roi}.

In this note, we study SUSY enhancement along an infinite class of RG flows starting from strongly interacting 4D $\CN=2$ SCFTs labeled by integers $(n,k)\ge(2,3)$\footnote{More precisely, as we will see below, these theories are specified by Young diagrams that are determined by $(n,k)$.} that do not have known Lagrangians\footnote{These theories lack $\CN=2$ Lagrangians because they have $\CN=2$ chiral operators of non-integer scaling dimension. Moreover, they do not have known UV Lagrangians in the sense of \cite{Gadde:2015xta,Maruyoshi:2016tqk,Maruyoshi:2016aim,Agarwal:2016pjo,Benvenuti:2017bpg,Giacomelli:2017ckh,Agarwal:2017roi}.} and ending at IR fixed points with thirty-two (Poincar\'e plus special) supercharges.\footnote{Several examples of four-dimensional flows from $\CN=2\to\CN=4$ were studied at the level of Coulomb branch geometries in \cite{Argyres:2016xmc}.} In particular, we provide evidence that these $\CN=2$ SCFTs flow, upon turning on \lq\lq mass terms"\footnote{More accurately, these are deformations of the superpotential by dimension two holomorphic moment maps in the same $\CN=2$ multiplets as certain flavor symmetries.} and compactifying the theories on an $S^1$ of radius $r$, to 3D $\CN=8$ SCFTs. These latter SCFTs can also be reached by turning on the gauge couplings of $u(n)$ 3D $\CN=8$ SYM for arbitrary $(n,k)\ge(2,3)$. In the case of $(n,k)=(2,3)$, we provide arguments that the $r\to\infty$ limit of the flow is to a 4D theory with $\CN=4$ SUSY.

While we believe it is likely that the $r\to\infty$ limits of these flows for any $(n,k)\ge(2,3)$ have 4D $\CN=4$ SUSY (with a $3n$ complex dimensional moduli space) in the IR, we leave a detailed study of this question and an analysis of the resulting spectra to future work \cite{toAppear}. One motivation for this note is simply to identify a space of theories in which SCFTs with $\CN=4$ SUSY in four dimensions may plausibly emerge somewhat more unconventionally. We hope these constructions will shed light on the space of possible $\CN=4$ theories (perhaps even on the question of whether these theories are necessarily of SYM type).

The plan of this paper is as follows. In the next section we describe how our UV theories are engineered starting from the $A_{N-1}$ $(2,0)$ theory. We then discuss the case of $(n,k)=(2,3)$ and motivate certain expectations for the corresponding RG flow from the superconformal index discussion of \cite{Buican:2017fiq}. We comment on the nature of the 4D IR fixed point that emerges in the $r\to\infty$ limit. Finally, we generalize our discussion to arbitrary $(n,k)\ge(2,3)$.

\newsec{The UV starting points}
Our particular UV 4D $\CN=2$ SCFTs are obtained from certain twisted compactifications of the $A_{N-1}$ 6D $(2,0)$ theory on a Riemann surface, $\CC=\mathbb{CP}^1$. A co-dimension two defect intersects $\CC$ at $z=\infty$ giving rise to an irregular puncture at this point \cite{Xie:2012hs} (see also \cite{Gaiotto:2009hg,Bonelli:2011aa}). In our class of theories, $\CC$ has no additional punctures.

One convenient way of studying certain aspects of the irregular puncture at $z=\infty$ and the resulting 4D theories is to first compactify the parent 6D theory on an $S^1$. We can then describe the irregular puncture in terms of the singular behavior of a twisted element of the vector multiplet of the corresponding $A_{N-1}$ 5D maximal SYM theory---the $\mathfrak{sl}(N,\mathbb{C})$-valued $(1,0)$-form, 
$\Phi_zdz$ (sometimes called the \lq\lq Higgs" field).\footnote{$\Phi_z$ is $\mathfrak{sl}(N,\mathbb{C})$-valued instead of $\mathfrak{su}(N,\mathbb{C})$-valued since it comes from $Y^1 + iY^2$ where $Y^1$ and $Y^2$ are two adjoint scalars in the 5D SYM. It is a $(1,0)$-form because of the twist.} Indeed, near the irregular puncture, we find \cite{Xie:2012hs,Gaiotto:2009hg}
\begin{equation}\label{Higgs}
\Phi_z = z^{\ell-2}T_{\ell-2}+z^{\ell-3}T_{\ell-3}+\cdots+T_{0}+{1\over z}T_{-1}+\cdots~,
\end{equation}
where the second set of ellipses contain non-singular terms in the limit $z\to\infty$, and the $T_i$ are traceless $N\times N$ matrices. In the above equation, $\ell>1$ is an integer (the case $\ell=1$ describes a regular singularity and is not relevant to our discussion below; the case $\ell\not\in\mathbb{Z}$ is also not relevant).

Combined with a gauge field on $\mathcal{C}$, the configuration in \eqref{Higgs} forms a solution to Hitchin's equations and describes the Higgs branch of the mirror of the $S^1$ reduction of our 4D theories of interest (the reduction of the 5D theory on $\CC$). Therefore, it describes the Coulomb branch of the direct $S^1$ reduction and also, via the base of the corresponding fibration, the Coulomb branch of the 4D theory itself. For example, the Seiberg-Witten curve of the 4D theory may be read off from the spectral curve \cite{Gaiotto:2009hg,Xie:2012hs}
\begin{equation}\label{SW}
{\rm det}(x-\Phi_z)=0~.
\end{equation}

In order for the description of the moduli space to not jump discontinuously as a function of the parameters residing in the $T_i$, a sufficient condition on the $T_i$ is that they are regular\footnote{Note that the puncture of $\CC$ is still irregular!} semisimple (see \cite{Witten:2007td} and references therein for a discussion in a closely related context). In particular, this statement means that the $T_i$ can be brought to the form of diagonal matrices with non-degenerate eigenvalues. These singularities give rise to 4D theories with Coulomb branch operators of non-integer scaling dimensions and generalize the theories described in \cite{Argyres:1995jj,Argyres:1995xn}.\footnote{Just as in the case of regular singularities, irregular singularities may be enriched by the presence of certain co-dimension one symmetry defects. Such a construction can lead to 4D SCFTs if there is also a regular singularity present \cite{Wang:2018gvb}.}

The above class of theories, while very broad, is (modulo some caveats we will discuss) not closed under the natural SCFT operation of conformal gauging \cite{Buican:2014hfa} or under the RG flow. In fact, these SCFTs form a part of a much broader but still relatively poorly understood class of theories called the \lq\lq type III" theories \cite{Xie:2012hs} (these theories are expected to exhibit various interesting phenomena; e.g., see \cite{Buican:2017fiq,Xie:2017vaf,Wang:2015mra,Xie:2017aqx}).

To define the type III SCFTs, we relax the condition of regularity of the $T_i$. In this case, the requirement of smoothness away from the origin of the moduli space implies that \cite{Witten:2007td}
\begin{equation}\label{Levi}
L_{-1}\subseteq L_0\subseteq\cdots\subseteq L_{\ell-2}~,
\end{equation}
where the $L_a$ are the Levi subalgebras associated with the $T_i$.\footnote{Specifically, $L_a$ is defined as the centralizer (in $A_{N-1}$) of the $T_i$ with $a\le i\le\ell-2$. Note that the conditions in \eqref{Levi} are necessary but not sufficient to have a sensible SCFT.} This restriction can be conveniently described in terms of certain Young diagrams \cite{Xie:2012hs}
\begin{equation}\label{Young}
T_i\ \leftrightarrow\ Y_i=[n_{i,1},n_{i,2},\cdots,n_{i,k_i}]~,\ \ \ n_{i,a}\ge n_{i,a+1}\in\mathbb{Z}_{>0}~,\ \ \ \sum_{a=1}^{k_i}n_{i,a}=N~,
\end{equation}
where the columns of height $n_{i,a}$ represent the eigenvalue degeneracies of the $T_i$. The condition \eqref{Levi} amounts to the statement that Young diagram $i$ and Young diagram $i-1$ are related by taking some number of columns (possibly zero) in diagram $i$ and decomposing each of them into columns in diagram $i-1$.

In this picture, the $T_{-1}$ matrix has a special status: it contains mass parameters (or, equivalently, vevs for the corresponding background vector multiplets) of the theory. By $\CN=2$ SUSY, such mass parameters correspond to elements of the Cartan subalgebra of the $\CN=2$ flavor symmetry group. In particular, we see that the rank of the flavor symmetry group, $G$, satisfies
\begin{equation}
{\rm rank}(G)\ge k_{-1}-1~,
\end{equation}
where the inequality is saturated for cases in which all symmetries are visible in the Hitchin system description (see the next section for an example with hidden symmetries).

As we will discuss below, one particular piece of progress in understanding type III theories relevant to us in this note is the first computation of the superconformal index in the non-regular case \cite{Buican:2017fiq}.\footnote{The associated chiral algebra, in the sense of \cite{Beem:2013sza}, was also determined in \cite{Buican:2017fiq}.}

In the remainder of this work, the particular theories we will be interested in have type III singularities of the form
\begin{equation}\label{class}
Y_{0,1}=[n,\cdots,n]~, \ \ \ Y_{-1}=[n,\cdots,n,n-1,1]~,
\end{equation}
where $n\ge2$, there are $k_{0,1}=k\ge3$ columns in $Y_{0,1}$, and there are $k_{-1}=k+1\ge4$ columns in $Y_{-1}$ (so that $N=nk$). We discuss the case of $(n,k)=(2,3)$ in the next section.

\newsec{The $(n,k)=(2,3)$ case}
In this section, we specialize to the UV $\CN=2$ SCFT given by the Young diagrams
\begin{equation}\label{23}
Y_1=Y_0=[2,2,2]~, \ \ \ Y_{-1}=[2,2,1,1]~.
\end{equation}
This theory was originally described implicitly in \cite{Xie:2012hs}. However, the construction in \cite{Buican:2014hfa} makes it clear that, although such a type III non-regular theory might seem exotic, it actually arises quite naturally when one uses more traditional SCFTs as building blocks. Indeed, the setup in \cite{Buican:2014hfa} starts by taking two copies of the isolated $(A_1, D_4)$ SCFT,\footnote{This theory was discovered in \cite{Argyres:1995xn} as a singular point on the Coulomb branch of $\CN=2$ $su(2)$ SQCD with $N_f=3$, but we follow the naming conventions of \cite{Cecotti:2010fi}.} adding nine hypermultiplets, and conformally gauging a diagonal $su(3)$ flavor symmetry.\footnote{Note that the resulting theory has $Y_{-1,0,1}=[2,2,1,1]$ and is non-regular type III even though the various isolated SCFT building blocks are not: the hypermultiplet is described by $Y_{-1,0,1}=[1,1]$, while $(A_1, D_4)$ is described by $Y_{-1,0,1}=[1,1,1]$. As alluded to above, this discussion shows (modulo potential dualities involving theories with one irregular singularity and a regular one) that the theories described by regular semisimple $T_i$ are not closed under conformal gauging.} Then, as one dials the $su(3)$ coupling to infinity, a dual weakly coupled description emerges with a diagonal $su(2)$ of an $(A_1, D_4)$ theory and the so-called $\CT_{3,{3\over2}}$ theory gauged. The $\CT_{3,{3\over2}}$ theory is another name for the SCFT with $Y_{0,1}=[2,2,2]$ and $Y_{-1}=[2,2,1,1]$.

The $\CT_{3,{3\over2}}$ theory has $su(2)^2\times su(3)$ flavor symmetry (of which a diagonal $su(2)\subset su(2)^2$ is gauged in the above duality), although only an $su(2)\times su(3)$ symmetry is visible in \eqref{23} (according to the analysis in \cite{Buican:2017fiq}, this theory splits into an interacting piece, $\CT_X$, with $su(2)\times su(3)$ flavor symmetry, and a free hypermultiplet with $su(2)$ flavor symmetry). More precisely, we see from \eqref{23} that $k_{-1}=4$, and so the visible flavor symmetry has rank three.

One remarkable feature of the duality described in \cite{Buican:2014hfa} is that, even though the theory in question is constructed from various strongly interacting non-Lagrangian building blocks (the $(A_1, D_4)$ and $\CT_{3,{3\over2}}$ SCFTs), each of these building blocks has certain observables that are closely related to the corresponding observables in free theories \cite{Buican:2017fiq}.\footnote{Heuristically this connection can be anticipated by noting that the $(A_1, D_4)$ theories play a role in the duality of \cite{Buican:2014hfa} that is reminiscent of the role played by hypermultiplets in the original duality of \cite{Argyres:2007cn}. In the case of the $(A_1, D_4)$ SCFT (and its generalizations), this connection was further explored in \cite{Buican:2017rya}.}

In the case of the $\CT_{3,{3\over2}}$ theory, the connection with free fields can be seen by examining its Schur index.\footnote{For an introduction to the Schur index, see \cite{Gadde:2011uv,Rastelli:2016tbz}.} After removing a decoupled free hypermultiplet to obtain the $\CT_X$ SCFT discussed above, we have the following Schur index \cite{Buican:2017fiq}
\begin{equation}\label{schurTX}
\CI=\sum_{\lambda=0}^{\infty}q^{{3\over2}\lambda}{\rm P.E.}\left[{2q^2\over1-q}+2q-2q^{1+\lambda}\right]{\rm ch}_{R_{\lambda}}^{su(2)}(q,w){\rm ch}_{R_{\lambda,\lambda}}^{su(3)}(q,z_1,z_2)~,
\end{equation}
where $\lambda$ is an integer, $q$ is a superconformal fugacity, and $w, z_{1,2}$ are flavor fugacities for $su(2)$ and $su(3)$ respectively. In \eqref{schurTX}, $ch_{R_{\lambda}}^{su(2)}$ and $ch_{R_{\lambda}}^{su(2)}$ are characters for modules of $\widehat{su(2)}_{-2}$ and $\widehat{su(3)}_{-3}$ affine Kac-Moody algebras at the crtitical level with primaries transforming with Dynkin labels $\lambda$ and $(\lambda,\lambda)$ of $su(2)$ and $su(3)$ respectively. Finally, \lq\lq P.E." stands for \lq\lq plethystic exponential" and is defined as follows
\begin{equation}
P.E.\left[f(x_1,\cdots,x_r)\right]\equiv\sum_{n=1}^{\infty}\exp\left({1\over n}f(x_1^n,\cdots,x_r^n)\right)~.
\end{equation}

The formula in \eqref{schurTX} is closely related to the index for 8 free half-hypermultiplets (the so-called $T_2$ theory \cite{Gaiotto:2009we})
\begin{equation}\label{schurT2}
\CI_{T_2}=\sum_{\lambda=0}^{\infty} q^{\lambda\over2}\ {\rm P.E.}\left[{2q^2\over1-q}+2q-2q^{1+\lambda}\right]{\rm ch}_{R_{\lambda}}^{su(2)}(q,x){\rm ch}_{R_{\lambda}}^{su(2)}(q,y){\rm ch}_{R_{\lambda}}^{su(2)}(q,z)~,
\end{equation}
where $x, y, z$ are fugacities for the $su(2)^3\subset sp(4)$ flavor symmetry (the particular re-writing of the $T_2$ index above was suggested in \cite{Lemos:2014lua}). Indeed, in both cases we sum over a \lq\lq diagonal" set of representations (of $su(2)^3$ in the $T_2$ case and of $su(2)\times su(3)$ in the $\CT_X$ case), and the structure constants (the plethystic exponential factors) are identical.

The $T_2$ theory has a natural connection with $su(2)$ $\CN=4$ SYM. Indeed, by diagonally gauging an $su(2)\times su(2)$ factor we are left with $su(2)$ $\CN=4$ SYM and a decoupled free hypermultiplet. Note that the remaining $\CN=2$ $su(2)$ flavor symmetry becomes part of the $su(4)_R$ symmetry of the $\CN=4$ theory.\footnote{Technically this is a diagonal flavor symmetry that acts both on the SYM theory and the decoupled hyper. Note that the $su(2)$ symmetry of the $\CN=4$ factor has a Witten anomaly \cite{Witten:1982fp}: it has 3 doublets charged under it (the corresponding holomorphic moment maps are $\sum_a Q^aQ^a, \sum_a\tilde Q^a\tilde Q^a, \sum_a\tilde Q^aQ^a$). This anomaly translates into the fact that, at generic points on the moduli space, we have a massless $u(1)$ $\CN=4$ theory: the singlet hypermultiplet is a doublet of $su(2)$ and therefore also gives rise to a Witten anomaly.} The deformation that connects $T_2$ to $\CN=4$ SYM is exactly marginal (although if we just want to get $\CN=4$, then we should also turn on a mass parameter for the hypermultiplet or else add a decoupled $u(1)$ $\CN=2$ vector multiplet).

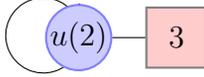
\begin{figure}
\begin{center}
\vskip .5cm
\begin{tikzpicture}[place/.style={circle,draw=blue!50,fill=blue!20,thick,inner sep=0pt,minimum size=6mm},transition2/.style={rectangle,draw=black!50,fill=red!20,thick,inner sep=0pt,minimum size=8mm},auto]
\draw [black] (3,.2) arc [radius=0.5, start angle=20, end angle= 340];
\node[place] (2) at (3,0) [shape=circle] {$u(2)$} edge [-] node[auto]{} (2);
\node[transition2] (3) at (4.3,0)  {$3$} edge [-] node[auto]{} (2);
\end{tikzpicture}
\caption{The quiver corresponding to the $S^1$ reduction of the $\CT_{3,{3\over2}}$ SCFT \cite{Buican:2017fiq}. Here the closed loop attached to the gauge node denotes an adjoint hypermultiplet of $u(2)$. This adjoint breaks up into a ${\bf3}+{\bf1}$ of $su(2)\subset u(2)$, with the singlet corresponding to the free decoupled hyper in $\CT_{3,{3\over2}}=\CT_X\oplus{\rm hyper}$ \cite{Buican:2017fiq}.}
\label{quiver0}
\end{center}
\end{figure}

\begin{figure}
\begin{center}
\vskip .5cm
\begin{tikzpicture}[place/.style={circle,draw=blue!50,fill=blue!20,thick,inner sep=0pt,minimum size=6mm},transition/.style={rectangle,draw=black!50,fill=black!20,thick,inner sep=0pt,minimum size=5mm},transition2/.style={rectangle,draw=black!50,fill=red!20,thick,inner sep=0pt,minimum size=8mm},auto]
\node[place] (2) at (.5,0) [shape=circle] {$u(2)$} edge [-] node[auto]{} (2);
\node[place] (3) at (2.,0) [shape=circle] {$u(2)$} edge [-] node[auto]{} (2);
\node[place] (4) at (1.25,-1) [shape=circle] {$u(2)$} edge[-] (2) edge[-] (3);
\node[transition2] (1) at (3.4,0) {$1$} edge[-] (3);
\end{tikzpicture}
\caption{The quiver corresponding to the mirror of the $S^1$ reduction of the $\CT_{3,{3\over2}}$ theory \cite{Buican:2017fiq,Xie:2012hs}. The Young diagrams describing the $\CT_{3,{3\over2}}$ theory are $Y_{1,0}=[2,2,2],~ Y_{-1}=[2,2,1,1]$ \cite{Xie:2012hs}.}
\label{quiver1}
\end{center}
\end{figure}
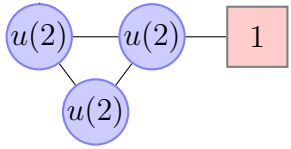

The $\CT_{3,{3\over2}}$ SCFT also has a connection to $\CN=4$. For example, as in the case of $su(2)$ $\CN=4$ SYM, the $su(2)\subset su(4)_R$ $\CN=2$ flavor symmetry of the interacting piece (the $\CT_X\subset\CT_{3,{3\over2}}$ theory) has a global Witten anomaly \cite{Buican:2017fiq}.\footnote{In the $su(2)$ $\CN=4$ case, this statement follows from the fact that the adjoint hypermultiplet transforms as three doublets of the $su(2)\subset su(4)_R$ symmetry. By similar reasoning, there is a non-vanishing Witten anomaly for this symmetry in $su(2r)$ $\CN=4$ theories.} More generally, it follows from anomaly matching that any $\CN=4$ theory (Lagrangian or not) with a rank one Coulomb branch (by which we mean that the low energy theory consists of a massless $U(1)$ $\CN=4$ vector multiplet at generic points along the three-real-dimensional moduli space) must have a non-vanishing Witten anomaly for the $su(2)\subset su(4)_R$ $\CN=2$ symmetry.\footnote{This statement generalizes for odd rank $\CN=4$ theories (again without appealing to the existence of a Lagrangian).}

Another connection between the $\CT_{3,{3\over2}}$ SCFT and $\CN=4$ can be found by, instead of introducing dynamical gauge fields (for $su(2)\times su(2)$) as in the $T_2$ case, introducing vevs for background gauge fields (i.e., mass terms) for the $su(3)$ symmetry. This statement is most obvious by first considering the $S^1$ reduction of the $\CT_{3,{3\over2}}$ theory. At the level of the index \eqref{schurTX}, this reduction is implemented by taking $q\to1$ (which corresponds to taking the radius of the $S^1\subset S^1\times S^3$ factor in the index to zero) and throwing away a flavor-independent divergent prefactor that encodes certain anomalies of the 4D theory (see \cite{DiPietro:2014bca,Buican:2015ina,Buican:2015hsa,Ardehali:2015bla,Buican:2017uka}). Performing this procedure, we showed in \cite{Buican:2017fiq} that \eqref{schurTX} reduces to the $S^3$ partition function of the 3D theory in Fig. \ref{quiver0}. This result confirms the rules conjectured in \cite{Xie:2012hs}, which produce the mirror quiver gauge theory in Fig. \ref{quiver1} (e.g., see \cite{Cremonesi:2014xha}).

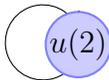
\begin{figure}
\begin{center}
\vskip .5cm
\begin{tikzpicture}[place/.style={circle,draw=blue!50,fill=blue!20,thick,inner sep=0pt,minimum size=6mm},transition2/.style={rectangle,draw=black!50,fill=red!20,thick,inner sep=0pt,minimum size=8mm},auto]
\draw [black] (3,.2) arc [radius=0.5, start angle=20, end angle= 340];
\node[place] (2) at (3,0) [shape=circle] {$u(2)$} edge [-] node[auto]{} (2);
\end{tikzpicture}
\caption{The quiver describing the endpoint of the flow initiated by turning on generic $su(3)$ mass parameters in Fig. \ref{quiver0}. We find a $u(2)$ $\CN=8$ theory (the $u(1)$ piece becomes a direct sum of a twisted hypermultiplet and a conventional hypermultiplet).}
\label{quiver2}
\end{center}
\end{figure}

From Fig. \ref{quiver0}, it is clear that if we turn on any superpotential mass term for the fundamental flavors we will flow to an $\CN=8$ SCFT that is the IR endpoint of the usual $\CN=8$ $u(2)$ SYM flow.\footnote{It is also clear that the Witten anomaly of the 4D $\CT_X$ theory is reflected in the fact that there are three doublets of the flavor $su(2)$ arising from the adjoint hypermultiplet of $su(2)\subset u(2)$.} This theory is then the same as the dimensional reduction of the $u(2)$ 4D $\CN=4$ theory.

\begin{figure}
\begin{center}
\vskip .5cm
\begin{tikzpicture}[place/.style={circle,draw=blue!50,fill=blue!20,thick,inner sep=0pt,minimum size=6mm},transition2/.style={rectangle,draw=black!50,fill=red!20,thick,inner sep=0pt,minimum size=8mm},auto]
\draw [black] (3,.2) arc [radius=0.5, start angle=20, end angle= 340];
\node[place] (2) at (3,0) [shape=circle] {$u(2)$} edge [-] node[auto]{} (2);
\node[transition2] (3) at (4.3,0)  {$1$} edge [-] node[auto]{} (2);
\end{tikzpicture}
\caption{The quiver corresponding to the IR endpoint of the RG flow from Fig. \ref{quiver0} after turning on masses for two fundamental flavors (these are non-generic $su(3)$ mass parameters in \eqref{3Dmasses}). This theory has accidental $\CN=8$ supersymmetry in the IR as in Fig. \ref{quiver2} \cite{Kapustin:2010xq}.}
\label{quiver3}
\end{center}
\end{figure}
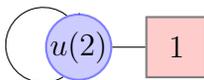

To see that we end up with 3D $\CN=8$ for any value of the superpotential mass terms, note that these mass terms are valued in the adjoint of $su(3)$ and can be parameterized as follows
\begin{equation}\label{3Dmasses}
m={\rm diag}( m_1,m_2, -m_1-m_2)~.
\end{equation}
Therefore, turning on generic $m_{1,2}$ results in giving masses to all the fundamental flavors, and we are left with the $\CN=8$ quiver in Fig. \ref{quiver2}. On the other hand, if we choose $m_1\ne0$ with $m_2=0$, $m_1=0$ with $m_2\ne0$, or $m_{1,2}\ne0$ with $m_1+m_2=0$, we give mass to two out of the three fundamental flavors and obtain the quiver in Fig. \ref{quiver3}. However, as is well-known (e.g., see \cite{Kapustin:2010xq}), this theory flows to the $\CN=8$ quiver of Fig. \ref{quiver2} in the IR.

\begin{figure}
\begin{center}
\includegraphics[height=1.8in,width=5.1in,angle=0]{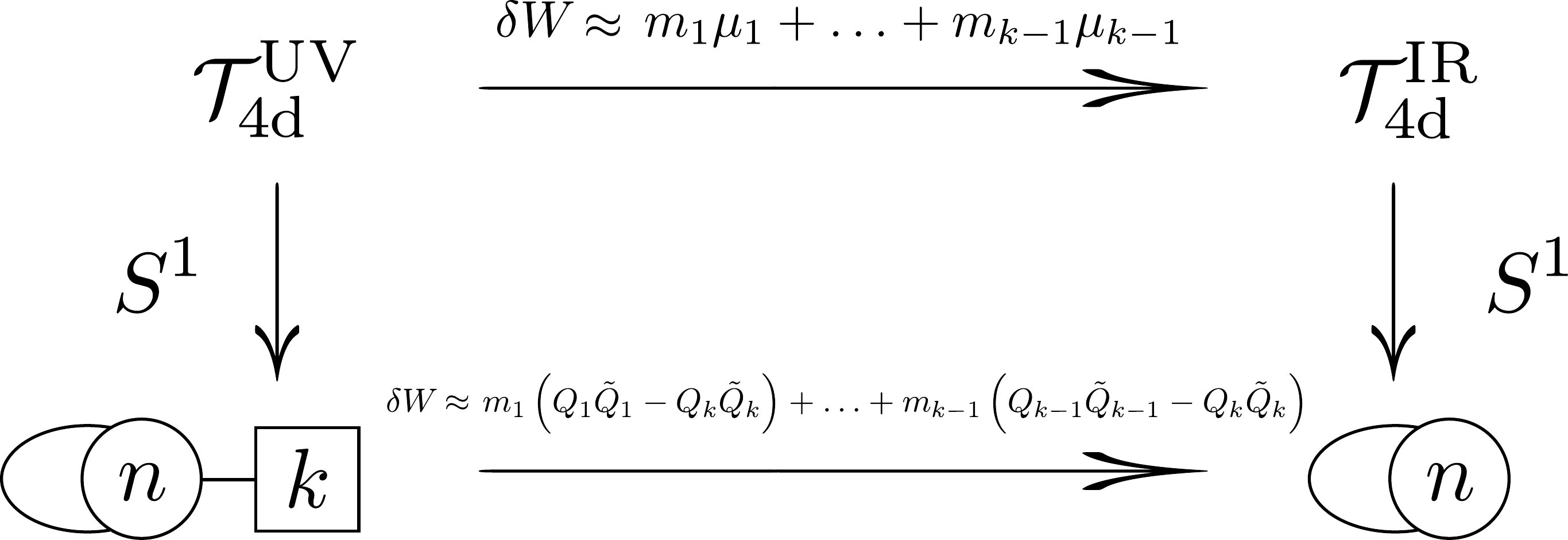}
\caption{The RG flows described in this note, with $\CT^{UV}_{4d}$ the non-Lagrangian 4D $\CN=2$ SCFT described around \eqref{class}. Horizontal arrows indicate superpotential deformations by holomorphic moment maps / mass terms of $su(k)$. Vertical arrows indicate $S^1$ reductions from 4D to 3D. All arrows preserve eight Poincar\'e supercharges. As described in the text, we expect this diagram to commute.}
\label{RGflow}
\end{center}
\end{figure}

Combining the procedure of putting the theory on a circle with turning on $su(3)$ mass terms gives us our desired RG flow from sixteen to thirty-two supercharges (see Fig. \ref{RGflow} with $(n,k)=(2,3)$). Indeed, this procedure is unambiguous since the 4D $su(3)$ holomorphic moment maps get mapped to gauge-invariant bilinears of the 3D theory
\begin{equation}
\mu_{1}\to_{r\to0} Q_1\tilde Q^1-Q_3\tilde Q^3~, \ \ \ \mu_{2}\to_{r\to0} Q_2\tilde Q^2-Q_3\tilde Q^3~,
\end{equation}
where $r$ is the $S^1$ radius, and $Q$, $\tilde Q$ are fundamental flavors of $u(2)$. In these expressions, gauge indices have been contracted, and the remaining indices are $su(3)$ flavor indices. Moreover, since there are no non-perturbative $\CN=4$-preserving deformations we can contemplate that arise from putting the theory on a circle,\footnote{This situation is unlike the one considered in \cite{Aharony:2013dha} for 4D $\CN=1$ theories. Note that in our case, flavor symmetries are all non-anomalous in both 4D and 3D due to the larger amount of SUSY.} and since our mass deformation does not induce Chern-Simons terms in 3D, we expect the limit of reducing the theory on a circle and turning on mass terms to commute.

\subsection{The $r\to\infty$ limit and (exotic?) 4D $\CN=4$}\label{N4}
Two natural questions arise from the above discussion:
\begin{itemize}
\item{Is the IR of the $r\to\infty$ limit of the above RG flow (i.e., $\CT^{IR}_{4d}$) a 4D $\CN=4$ SCFT?}
\item{If $\CT^{IR}_{4d}$ has 4D $\CN=4$ SUSY, is this theory $u(2)$ SYM?}
\end{itemize}
\noindent
Note that the presence of a 3D Lagrangian does not immediately shed light on these questions since, in principle, it is possible that the $\CN=8$ SUSY is accidental in 3D. Moreover, the existence of a 3D Lagrangian does not obviously imply a 4D Lagrangian. Indeed, the $\CT_{3,{3\over2}}$ theory does not have a Lagrangian even though the $S^1$ reduction does (as in Fig. \ref{quiver0}).

One way to explore these questions is to construct the Seiberg-Witten curve for the $\CT_{3,{3\over2}}$ theory\footnote{By \eqref{SW}, the Seiberg-Witten curve for this theory is guaranteed to exist. In general, it is not clear whether a given $\CN=2$ SCFT must have such a curve.} and define a scaling limit that produces the Seiberg-Witten curve of the IR theory, $\CT_{4d}^{IR}$ (e.g., see \cite{Gaiotto:2010jf} for a successful recent application of this technique). 

The Seiberg-Witten curve corresponding to an SCFT describes the Coulomb branch that one obtains by deforming the SCFT by relevant or marginal prepotential couplings, mass parameters (i.e., background vector multiplets), and expectation values of $\CN=2$ chiral operators. In general it is not clear whether a particular marginal or relevant parameter of the UV SCFT must necessarily appear in the curve, since the curve is an effective description of the theory.\footnote{For example, in the case of the $\CT_{3,{3\over2}}$ theory, there are actually two independent $su(2)$ mass parameters (since we have flavor symmetry $su(2)^2\times su(3)$), but, as discussed in \cite{Buican:2017fiq}, only one appears in the curve coming from \eqref{SW}. Note that this additional mass parameter might become visible through an alternate construction of the curve that does not go through the particular Hitchin system we described above.} However, all parameters appearing in the curve are of the type just described.

To obtain the curve in the case of the $\CT_{3,{3\over2}}$ theory, we start by writing the Higgs field as in \eqref{Higgs}
\begin{eqnarray}\label{Higgs23}
\Phi_z&=&z\ {\rm diag}(a_1,a_1,a_2,a_2,-a_1-a_2,-a_1-a_2)+{\rm diag}(b_1,b_1,b_2,b_2,-b_1-b_2,-b_1-b_2)\cr&+&{1\over z}\ {\rm diag}(m_1,m_1,m_2,m_2,-m_1-m_2+m_3,-m_1-m_2-m_3)\cr&+&{1\over z^2}\ {\rm diag}(c_1,c_2,c_3,c_4,c_5,-c_1-c_2-c_3-c_4-c_5)+\CO(z^{-3})~,
\end{eqnarray}
where the degeneracies of the eigenvalues in each singular term correspond to the Young diagrams in \eqref{23} (the non-singular pieces, starting with the $c_i$, describe vevs of $\CN=2$ chiral operators). In principle, since we are interested in studying the RG flow along the top arrow in Fig. \ref{RGflow}, we may turn off the $su(2)$ mass parameter.
At the level of \eqref{Higgs23}, this manoeuvre corresponds to setting $m_3=0$. 
Indeed, we then see that the singular sterms in \eqref{Higgs23} are subject to a natural action of the $S_3$ Weyl group of $su(3)$, which acts via permutation of the degenerate two-by-two blocks.\footnote{In particular, the curve we get from \eqref{SW} will be invariant under this action.} More precisely, this action corresponds to the simultaneous $S_3$ action on the 3-tuples
\begin{equation}
(a_1, a_2, a')~, \ \ \ (b_1, b_2, b')~, \ \ \ (m_1, m_2, m')~,
\end{equation}
where $a'=-a_1-a_2$, $b'=-b_1-b_2$, and $m'=-m_1-m_2$. In what follows, we will set $m_3=0$.

To write the curve for the $\CT_{3,{3\over2}}$ SCFT, it is convenient to first shift $x$ and $z$ by constants\footnote{Note that these shifts affect the 1-form only by exact terms, and BPS masses are unchanged.} so that the $\CO(z^0)$ matrix in \eqref{Higgs23} is of the form ${\rm diag}(0,0,0,0,b,b)$. Now, plugging this result into \eqref{SW} yields
\begin{align}\label{SW23}
u_2&+\Big((x-a_1z)(x-a_2z)(x+(a_1+a_2)z)+{M_1\over2}(x-a_1z)+{M_2\over2}(x-a_2z)\cr&-b(x-a_1z)(x-a_2z)\Big)^2 +u_1\Big(-b(a_1-a_2)(x-a_1 z)(x-a_2z)\cr&-(x-a_1z)^2(x-a_2z)(a_1+2a_2)+(x-a_1z)(x-a_2z)^2(2a_1+a_2)\cr&+{a_1-a_2\over2}(M_1(x-a_1z)+M_2(x-a_2z))\Big)=0~,
\end{align}
where
\begin{align}
M_1=&-2(a_1+2a_2)m_2~,\ \ \ M_2=-2(2a_1+a_2)m_1~, \ \ \ u_1 =-(2a_1+a_2)(c_1+c_2)+2bm_1~, \cr u_2=&(a_1-a_2)^2((2a_1+a_2)c_1-bm_1)((2a_1+a_2)c_2-bm_1)~.
\end{align}
In the above equations, $u_1$ is the vev of the $\CN=2$ chiral ring generator of dimension $3/2$, while $u_2$ is the vev of the $\CN=2$ chiral ring generator of dimension $3$. The parameter $b$ is the relevant coupling of dimension $1/2$. The dimensionless parameters, $a_i$, are not physical since they are absorbed by changing coordinates, $x$ and $z$. The above curve transforms homogenously (with the couplings and vevs acting as spurions) under the $u(1)_R$ scaling of the UV SCFT.

To study the RG flow described by the top arrow in Fig. \ref{RGflow}, we would like to turn on some RG scale, $m$, in \eqref{SW23} and take $m\to\infty$.\footnote{Note that this method is a rather indirect way of studying the RG flow: we try to carve out the Coulomb branch of $\CT^{IR}_{4d}$ as a subspace of the Coulomb branch of $\CT^{UV}_{4d}$ rather than considering the flow starting from the UV SCFT and then deforming by $\delta W\sim m_1\mu_1+m_2\mu_2$ with zero vevs (the 3D picture of the RG flow suggests that we should remain at the origin of the 4D Coulomb branch).} We can make contact with a curve resembling that of the $su(2)$ $\CN=4$ theory (we expect to have an additional $\CN=4$ $u(1)$ decoupled) if we set
\begin{align}\label{scalinglim}
u_1&=0~, \ \ \ M_1=m~, \ \ \ M_2=0~, \ \ \ b=qm^{1\over2}~, \cr x-a_1z&=2m^{-{1\over2}}X~, \ \ \ x-a_2z=m^{1\over2}Z~, \ \ \ u_2=-Um~.
\end{align}
Here, $X$ and $Z$ act as good coordinates describing the curve of $\CT^{IR}_{4d}$ and do not scale with $m$ (they have scaling dimension one and zero respectively) while $U$ is a dimension two vev.  We interpret some of the terms that are turned off as decoupling or becoming irrelevant along the RG flow (although, as discussed below, additional quantities decouple). 
 Keeping only the leading terms in \eqref{SW23} as $m\to\infty$ and solving for $X$, we obtain
\begin{equation}\label{N4curve}
X^2={U\over\left({2(2a_1+a_2)\over a_1-a_2}Z^2-2qZ+1\right)^2}~.
\end{equation}
This equation is the $su(2)$ $\CN=4$ curve\footnote{By \eqref{scalinglim}, the 1-form is (modulo exact terms) also the 1-form for $su(2)$ $\CN=4$ (up to a constant we can tune).} tuned to a cusp (i.e., a weak gauge coupling limit). Indeed, although there is an apparently dimension zero parameter, $q$, arising in \eqref{scalinglim} (reflecting the fact that the dimension half coupling in the Hitchin system forms a dimensionless combination with the square-root of the mass parameter), this naively marginal parameter is irrelevant in the IR description given above.

Note that we may also exchange $a_1\leftrightarrow a_2$ and $M_1\leftrightarrow M_2$ and obtain a similar limit of the curve. Finally, we may also construct a closely related limit of the curve by taking the linear combination $(2a_1+a_2)M_1+(a_1+2a_2)M_2$ to vanish.\footnote{Said more invariantly, when we take the scaling limit in \eqref{scalinglim}, ${\rm Tr}\ T_{-1}^2\to\infty$ and ${\rm Tr}\ T_{-1}^3\to0$.}

We have not been able to find a more general non-trivial scaling limit than the one described above. In particular, we are not able to see the putative $\CT^{IR}_{4d}$ marginal deformation (away from the cusp) in the Seiberg-Witten description we have found. Note that if $\CT^{IR}_{4d}$ has $\CN=4$ SUSY, it necessarily possesses an exactly marginal deformation residing in the stress-energy tensor multiplet (this statement follows from $\CN=4$ SUSY and is not related to the existence of an $\CN=4$ Lagrangian). However, there may be several reasons for the absence of this marginal direction in the Coulomb branch effective action:
\begin{itemize}
\item A radical option is that $\CT_{4d}^{IR}$ is an exotic 4D $\CN=4$ theory without a Lagrangian description. In a Lagrangian theory, we expect that W-boson masses will vary as a function of the marginal gauge coupling (this statement follows from the Higgs mechanism). These changes in mass are reflected in the periods of the curve, since the W-bosons are BPS particles. On the other hand, as far as we are aware, there is no argument that the most general exactly marginal parameter in an $\CN=2$ or $\CN=4$ SCFT must appear in the IR effective action captured by the Seiberg-Witten description. If the IR effective action is indeed given by \eqref{N4curve}, it means that the exactly marginal parameter of the UV SCFT becomes irrelevant in the IR (as opposed to being related to a marginal coupling in the IR). In this case, varying the exactly marginal parameter may have a more profound effect on the non-BPS sector.
\item A less radical option is that the exactly marginal parameter of $\CT_{4d}^{IR}$ is a standard gauge coupling, but it is hidden in the flow from $\CT_{3,{3\over2}}$. This option is not implausible since the existence of a conformal manifold is accidental in this case. If this possibility is realized, then perhaps the marginal coupling becomes visible by choosing a different UV starting point than $\CT_{4d}^{UV}$.
\item The most conservative option is simply that there is a more general scaling limit that describes the curve of $\CT^{IR}_{4d}$ for all values of the exactly marginal parameter. In this case,  $\CT_{4d}^{IR}$ may again be a standard $\CN=4$ Lagrangian theory.
\end{itemize}
We hope to conduct a more detailed study of these options using additional techniques \cite{toAppear}. In the next section we set this goal aside for now and present infinitely many generalizations of the above discussion.

\newsec{Generalizations}
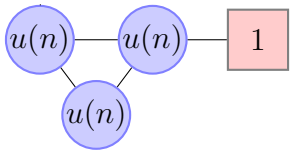
\begin{figure}
\begin{center}
\vskip .5cm
\begin{tikzpicture}[place/.style={circle,draw=blue!50,fill=blue!20,thick,inner sep=0pt,minimum size=6mm},transition/.style={rectangle,draw=black!50,fill=black!20,thick,inner sep=0pt,minimum size=5mm},transition2/.style={rectangle,draw=black!50,fill=red!20,thick,inner sep=0pt,minimum size=8mm},auto]
\node[place] (2) at (.5,0) [shape=circle] {$u(n)$} edge [-] node[auto]{} (2);
\node[place] (3) at (2.,0) [shape=circle] {$u(n)$} edge [-] node[auto]{} (2);
\node[place] (4) at (1.25,-1) [shape=circle] {$u(n)$} edge[-] (2) edge[-] (3);
\node[transition2] (1) at (3.4,0) {$1$} edge[-] (3);
\end{tikzpicture}
\caption{The quiver corresponding to the mirror of the $S^1$ reduction of the type III AD theory with $Y_{1,0}=[n,n,n],~ Y_{-1}=[n,n,n-1,1]$ \cite{Xie:2012hs}.}
\label{quiver4}
\end{center}
\end{figure}
It is rather straightforward to generalize the above discussion to other values of $n$ and $k$. For example, we can take any $n\ge2$. $\CT_{4d}^{UV}$ is now described by the following Young diagrams, which generalize \eqref{23}
\begin{equation}
Y_1=Y_0=[n,n,n]~, \ \ \ Y_{-1}=[n,n,n-1,1]~.
\end{equation}
Applying the discussion in \cite{Xie:2012hs}, one can easily check that this theory has rank $n$ with $\CN=2$ chiral ring generators of scaling dimensions
\begin{equation}\label{dims}
\Delta=\left\{{3\over2}~,\ 3~,\ {9\over2}~,\ \cdots~,\ {3n\over2}\right\}~.
\end{equation}
Note that, as in $\CN=4$, the scaling dimensions of chiral operators are integer multiples of the dimension of the lowest dimensional chiral operator (although here, unlike in $\CN=4$, the scaling dimension of the lowest dimensional chiral operator is half-integer).\footnote{In fact, the scaling dimensions of the operators in \eqref{dims} correspond to those of $u(n)$ $\CN=4$ SYM up to an overall multiplication by $3/2$.}

\begin{figure}
\begin{center}
\vskip .5cm
\begin{tikzpicture}[place/.style={circle,draw=blue!50,fill=blue!20,thick,inner sep=0pt,minimum size=6mm},transition2/.style={rectangle,draw=black!50,fill=red!20,thick,inner sep=0pt,minimum size=8mm},auto]
\draw [black] (3,.2) arc [radius=0.5, start angle=20, end angle= 340];
\node[place] (2) at (3,0) [shape=circle] {$u(n)$} edge [-] node[auto]{} (2);
\node[transition2] (3) at (4.3,0)  {$3$} edge [-] node[auto]{} (2);
\end{tikzpicture}
\caption{The quiver corresponding to the $S^1$ reduction of the type III AD theory with $Y_{1,0}=[n,n,n],~ Y_{-1}=[n,n,n-1,1]$. The closed loop attached to the gauge node denotes an adjoint hypermultiplet of $u(n)$.}
\label{quiver5}
\end{center}
\end{figure}
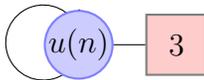

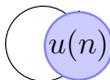
\begin{figure}
\begin{center}
\vskip .5cm
\begin{tikzpicture}[place/.style={circle,draw=blue!50,fill=blue!20,thick,inner sep=0pt,minimum size=6mm},transition2/.style={rectangle,draw=black!50,fill=red!20,thick,inner sep=0pt,minimum size=8mm},auto]
\draw [black] (3,.2) arc [radius=0.5, start angle=20, end angle= 340];
\node[place] (2) at (3,0) [shape=circle] {$u(n)$} edge [-] node[auto]{} (2);
\end{tikzpicture}
\caption{The result of turning on generic $su(3)$ masses in the quiver in Fig. \ref{quiver5}.}
\label{quiver6}
\end{center}
\end{figure}

In this case, the 3D mirror quiver generalizing Fig. \ref{quiver1} is given in Fig. \ref{quiver4} following the rules in \cite{Xie:2012hs}. The mirror of this quiver (i.e., the direct $S^1$ reduction) is the $u(n)$ theory with an adjoint hypermultiplet and three fundamental flavors as in Fig. \ref{quiver5} (e.g., see the discussion in \cite{Cremonesi:2014xha}).

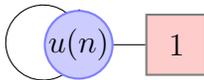
\begin{figure}
\begin{center}
\vskip .5cm
\begin{tikzpicture}[place/.style={circle,draw=blue!50,fill=blue!20,thick,inner sep=0pt,minimum size=6mm},transition2/.style={rectangle,draw=black!50,fill=red!20,thick,inner sep=0pt,minimum size=8mm},auto]
\draw [black] (3,.2) arc [radius=0.5, start angle=20, end angle= 340];
\node[place] (2) at (3,0) [shape=circle] {$u(n)$} edge [-] node[auto]{} (2);
\node[transition2] (3) at (4.3,0)  {$1$} edge [-] node[auto]{} (2);
\end{tikzpicture}
\caption{The result of turning on masses for two out of the three flavors in Fig. \ref{quiver6}. Quantum mechanically, this remaining fundamental flavor also gets a mass \cite{Kapustin:2010xq}.}
\label{quiver7}
\end{center}
\end{figure}

We may then reproduce the discussion for $n=2$ for general $n\ge2$ by turning on masses for the three fundamental flavors in the $S^1$ reduction. For generic masses, we end up with the quiver in Fig. \ref{quiver6}. For non-generic $su(3)$ masses, we end up with the quiver in Fig. \ref{quiver7}, which, by the discussion in \cite{Kapustin:2010xq}, flows to the 3D $\CN=8$ quiver in Fig. \ref{quiver6}. By combining the procedure of $S^1$ reduction with turning on masses, we again, as in the more detailed discussion of the $n=2$ case, get the commuting RG diagram of Fig. \ref{RGflow} with accidental enhancement to thirty-two (Poincar\'e plus special) supercharges in the IR. We again suspect (but have not proven) that the $r\to\infty$ limit of this flow has $\CN=4$ SUSY.

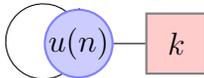
\begin{figure}
\begin{center}
\vskip .5cm
\begin{tikzpicture}[place/.style={circle,draw=blue!50,fill=blue!20,thick,inner sep=0pt,minimum size=6mm},transition2/.style={rectangle,draw=black!50,fill=red!20,thick,inner sep=0pt,minimum size=8mm},auto]
\draw [black] (3,.2) arc [radius=0.5, start angle=20, end angle= 340];
\node[place] (2) at (3,0) [shape=circle] {$u(n)$} edge [-] node[auto]{} (2);
\node[transition2] (3) at (4.3,0)  {$k$} edge [-] node[auto]{} (2);
\end{tikzpicture}
\caption{The quiver corresponding to the $S^1$ reduction of the type III AD theory with Young diagrams described in \eqref{general}. The closed loop attached to the gauge node denotes an adjoint hypermultiplet of $u(n)$.}
\label{quiver8}
\end{center}
\end{figure}

Finally, note that we can have an even more general UV starting point given by
\begin{equation}\label{general}
Y_{1,0}=[n,n,\cdots,n]~, \ \ \ Y_{-1}=[n,\cdots,n,n-1,1]~,
\end{equation}
where, as in \eqref{class}, $n\ge2$, there are $k\ge3$ columns in $Y_{0,1}$, and there are $k+1\ge4$ columns in $Y_{-1}$ (so that $N=nk$, where we obtain our theory from the $A_{N-1}$ $(2,0)$ theory). Here the mirror looks as in Fig. \ref{quiver4}, but now there is a $k$-sided polygon of $u(n)$ nodes with one node coupled to a fundamental flavor. The direct reduction of the theory is given in Fig. \ref{quiver8}. Just as in the previous cases, we may give masses to these $k$ fundamental flavors and flow to a theory with thirty-two (Poincar\'e plus special) supercharges, thus obtaining the RG diagram in Fig. \ref{RGflow}.

\newsec{Conclusions}
In this note we have studied an infinite set of RG flows that start from 4D $\CN=2$ SCFTs that lack a Lagrangian description and end up, after turning on generalized mass terms, flowing to theories that have thirty-two (Poincar\'e plus special) supercharges. We are able to demonstrate this fact compellingly when we also compactify these theories on a circle (and we have the flow diagram in Fig. \ref{RGflow}).

We also gave some preliminary, but far from conclusive, arguments that these theories flow to 4D $\CN=4$ SCFTs (at least for $(n,k)=(2,3)$) when we take the radius of the circle to infinity. One important matching quantity was the Witten anomaly for the $\CT_X\subset\CT_{3,{3\over2}}$ SCFT. In \cite{Buican:2017fiq}, we wondered how to construct such Witten anomalous theories directly in terms of punctured compactifications of the $(2,0)$ theory. Recently, there has been progress on this topic \cite{Tachikawa:2018rgw,Wang:2018gvb}. Moreover, the authors of \cite{Wang:2018gvb} find an $\CN=4$ theory starting directly from an irregular singularity (and a regular singularity, both in the presence of a co-dimension one defect). It would be interesting to see if their theory is related to $\CT_{4d}^{IR}$ in the case of $(n,k)=(2,3)$.

We have also seen that the scaling limit we chose does not reproduce the standard $\CN=4$ curve, since the IR description seems tuned to a cusp. As discussed in section \ref{N4}, this result may have various causes ranging from the existence of an exotic $\CN=4$ non-Lagrangian theory to the presence of a hidden marginal direction or to the existence of a more general scaling limit that describes the curve of $\CT_{4d}^{IR}$. It would be interesting to find out which of these options is realized \cite{toAppear}.\footnote{It would also be interesting to see if our RG flows shed any light on the question of classifying $su(N)$ $\CN=4$ SYM theories \cite{Aharony:2013hda,Argyres:2016yzz,Inaki} and if we can understand some of our theories from a holographic point of view (along the lines of \cite{Lin:2004nb}).}

\bigskip\bigskip

\ack{We are grateful to L.~Hollands for collaboration on related work and to S.~Giacomelli for various discussions. M.B. would like to thank the Galileo Galilei Institute for a stimulating environment during the workshop on \lq\lq Supersymmetric Quantum Field Theories in the Non-Perturbative Regime," where part of this work was completed. M. B.'s research is partially supported by the Royal Society under the grant \lq\lq New Constraints and Phenomena in Quantum Field Theory." Z. L. is supported by a Queen Mary University of London PhD studentship. T. N.'s research is partially supported by JSPS KAKENHI Grant Number 18K13547.}

\newpage
\bibliography{chetdocbib}

\begin{thebibliography}{10}
\ifx\href\asklfhas\newcommand{\href}[2]{#2}\fi
\ifx\arxivref\asklfhas\newcommand{\arxivref}[2]{\href{http://arxiv.org/abs/#1}{#2}}\fi
\ifx\doiref\asklfhas\newcommand{\doiref}[2]{\href{http://dx.doi.org/#1}{#2}}\fi
\parskip 0pt
\normalsize

\bibitem{Abel:2011wv}
S.~Abel, M.~Buican \& Z.~Komargodski,
\textit{``{Mapping Anomalous Currents in Supersymmetric Dualities}''},
\doiref{10.1103/PhysRevD.84.045005}{Phys.~Rev. \textbf{D84}, 045005
  (2011)\ignorespaces}\ignorespaces,
\normalsize{\texttt{\arxivref{1105.2885}{arXiv:1105.2885}}}\ignorespaces
\bibitem{Buican:2011ty}
M.~Buican,
\textit{``{A Conjectured Bound on Accidental Symmetries}''},
\doiref{10.1103/PhysRevD.85.025020}{Phys.~Rev. \textbf{D85}, 025020
  (2012)\ignorespaces}\ignorespaces,
\normalsize{\texttt{\arxivref{1109.3279}{arXiv:1109.3279}}}\ignorespaces
\bibitem{Buican:2012ec}
M.~Buican,
\textit{``{Non-Perturbative Constraints on Light Sparticles from Properties of
  the RG Flow}''},
\doiref{10.1007/JHEP10(2014)026}{JHEP \textbf{1410}, 026
  (2014)\ignorespaces}\ignorespaces,
\normalsize{\texttt{\arxivref{1206.3033}{arXiv:1206.3033}}}\ignorespaces
\bibitem{Collins:2016icw}
T.~C. Collins, D.~Xie \& S.-T. Yau,
\textit{``{K stability and stability of chiral ring}''},
\normalsize{\texttt{\arxivref{1606.09260}{arXiv:1606.09260}}}\ignorespaces
\bibitem{Cordova:2018cvg}
C.~C\'ordova, T.~T. Dumitrescu \& K.~Intriligator,
\textit{``{Exploring 2-Group Global Symmetries}''},
\normalsize{\texttt{\arxivref{1802.04790}{arXiv:1802.04790}}}\ignorespaces
\bibitem{Balents:2008vd}
L.~Balents, M.~P.~A. Fisher \& C.~Nayak,
\textit{``{Nodal Liquid Theory of the Pseudo-Gap Phase of High-T(c)
  Superconductors}''},
\doiref{10.1142/S0217979298000570}{Int.~J.~Mod.~Phys. \textbf{B12}, 1033
  (1998)\ignorespaces}\ignorespaces,
\normalsize{\texttt{\arxivref{cond-mat/9803086}{cond-mat/9803086}}}\ignorespaces
\bibitem{Grover:2013rc}
T.~Grover, D.~N. Sheng \& A.~Vishwanath,
\textit{``{Emergent Space-Time Supersymmetry at the Boundary of a Topological
  Phase}''},
\doiref{10.1126/science.1248253}{Science \textbf{344}, 280
  (2014)\ignorespaces}\ignorespaces,
\normalsize{\texttt{\arxivref{1301.7449}{arXiv:1301.7449}}}\ignorespaces
\bibitem{Lee:2006if}
S.-S. Lee,
\textit{``{Emergence of supersymmetry at a critical point of a lattice
  model}''},
\doiref{10.1103/PhysRevB.76.075103}{Phys.~Rev. \textbf{B76}, 075103
  (2007)\ignorespaces}\ignorespaces,
\normalsize{\texttt{\arxivref{cond-mat/0611658}{cond-mat/0611658}}}\ignorespaces
\bibitem{Aharony:2008ug}
O.~Aharony, O.~Bergman, D.~L. Jafferis \& J.~Maldacena,
\textit{``{N=6 superconformal Chern-Simons-matter theories, M2-branes and their
  gravity duals}''},
\doiref{10.1088/1126-6708/2008/10/091}{JHEP \textbf{0810}, 091
  (2008)\ignorespaces}\ignorespaces,
\normalsize{\texttt{\arxivref{0806.1218}{arXiv:0806.1218}}}\ignorespaces
\bibitem{Gaiotto:2018yjh}
D.~Gaiotto, Z.~Komargodski \& J.~Wu,
\textit{``{Curious Aspects of Three-Dimensional ${\cal N}=1$ SCFTs}''},
\normalsize{\texttt{\arxivref{1804.02018}{arXiv:1804.02018}}}\ignorespaces
\bibitem{Benini:2018bhk}
F.~Benini \& S.~Benvenuti,
\textit{``{$N=1$ QED in 2+1 dimensions: Dualities and enhanced symmetries}''},
\normalsize{\texttt{\arxivref{1804.05707}{arXiv:1804.05707}}}\ignorespaces
\bibitem{Gang:2018huc}
D.~Gang \& M.~Yamazaki,
\textit{``{An appetizer for supersymmetry enhancement}''},
\normalsize{\texttt{\arxivref{1806.07714}{arXiv:1806.07714}}}\ignorespaces
\bibitem{Gadde:2015xta}
A.~Gadde, S.~S. Razamat \& B.~Willett,
\textit{``{"Lagrangian" for a Non-Lagrangian Field Theory with $\mathcal N=2$
  Supersymmetry}''},
\doiref{10.1103/PhysRevLett.115.171604}{Phys.~Rev.~Lett. \textbf{115}, 171604
  (2015)\ignorespaces}\ignorespaces,
\normalsize{\texttt{\arxivref{1505.05834}{arXiv:1505.05834}}}\ignorespaces
\bibitem{Maruyoshi:2016tqk}
K.~Maruyoshi \& J.~Song,
\textit{``{Enhancement of Supersymmetry via Renormalization Group Flow and the
  Superconformal Index}''},
\doiref{10.1103/PhysRevLett.118.151602}{Phys.~Rev.~Lett. \textbf{118}, 151602
  (2017)\ignorespaces}\ignorespaces,
\normalsize{\texttt{\arxivref{1606.05632}{arXiv:1606.05632}}}\ignorespaces
\bibitem{Maruyoshi:2016aim}
K.~Maruyoshi \& J.~Song,
\textit{``{$ \mathcal{N}=1 $ deformations and RG flows of $ \mathcal{N}=2 $
  SCFTs}''},
\doiref{10.1007/JHEP02(2017)075}{JHEP \textbf{1702}, 075
  (2017)\ignorespaces}\ignorespaces,
\normalsize{\texttt{\arxivref{1607.04281}{arXiv:1607.04281}}}\ignorespaces
\bibitem{Agarwal:2016pjo}
P.~Agarwal, K.~Maruyoshi \& J.~Song,
\textit{``{$ \mathcal{N} $ =1 Deformations and RG flows of $ \mathcal{N} $ =2
  SCFTs, part II: non-principal deformations}''},
\doiref{10.1007/JHEP12(2016)103}{JHEP \textbf{1612}, 103
  (2016)\ignorespaces}\ignorespaces,
\normalsize{\texttt{\arxivref{1610.05311}{arXiv:1610.05311}}}\ignorespaces
\bibitem{Benvenuti:2017bpg}
S.~Benvenuti \& S.~Giacomelli,
\textit{``{Lagrangians for generalized Argyres-Douglas theories}''},
\doiref{10.1007/JHEP10(2017)106}{JHEP \textbf{1710}, 106
  (2017)\ignorespaces}\ignorespaces,
\normalsize{\texttt{\arxivref{1707.05113}{arXiv:1707.05113}}}\ignorespaces
\bibitem{Giacomelli:2017ckh}
S.~Giacomelli,
\textit{``{RG flows with supersymmetry enhancement and geometric
  engineering}''},
\normalsize{\texttt{\arxivref{1710.06469}{arXiv:1710.06469}}}\ignorespaces
\bibitem{Agarwal:2017roi}
P.~Agarwal, A.~Sciarappa \& J.~Song,
\textit{``{$ \mathcal{N} $ =1 Lagrangians for generalized Argyres-Douglas
  theories}''},
\doiref{10.1007/JHEP10(2017)211}{JHEP \textbf{1710}, 211
  (2017)\ignorespaces}\ignorespaces,
\normalsize{\texttt{\arxivref{1707.04751}{arXiv:1707.04751}}}\ignorespaces
\bibitem{Argyres:2016xmc}
P.~Argyres, M.~Lotito, Y.~Lu \& M.~Martone,
\textit{``{Geometric constraints on the space of $\mathcal{N}$ = 2 SCFTs. Part
  III: enhanced Coulomb branches and central charges}''},
\doiref{10.1007/JHEP02(2018)003}{JHEP \textbf{1802}, 003
  (2018)\ignorespaces}\ignorespaces,
\normalsize{\texttt{\arxivref{1609.04404}{arXiv:1609.04404}}}\ignorespaces
\bibitem{toAppear}
M.~Buican, L.~Hollands, Z.~Laczko \& T.~Nishinaka,
\textit{``In progress''}
\bibitem{Buican:2017fiq}
M.~Buican, Z.~Laczko \& T.~Nishinaka,
\textit{``{$ \mathcal{N} $ = 2 S-duality revisited}''},
\doiref{10.1007/JHEP09(2017)087}{JHEP \textbf{1709}, 087
  (2017)\ignorespaces}\ignorespaces,
\normalsize{\texttt{\arxivref{1706.03797}{arXiv:1706.03797}}}\ignorespaces
\bibitem{Xie:2012hs}
D.~Xie,
\textit{``{General Argyres-Douglas Theory}''},
\doiref{10.1007/JHEP01(2013)100}{JHEP \textbf{1301}, 100
  (2013)\ignorespaces}\ignorespaces,
\normalsize{\texttt{\arxivref{1204.2270}{arXiv:1204.2270}}}\ignorespaces
\bibitem{Gaiotto:2009hg}
D.~Gaiotto, G.~W. Moore \& A.~Neitzke,
\textit{``{Wall-crossing, Hitchin Systems, and the WKB Approximation}''},
\normalsize{\texttt{\arxivref{0907.3987}{arXiv:0907.3987}}}\ignorespaces
\bibitem{Bonelli:2011aa}
G.~Bonelli, K.~Maruyoshi \& A.~Tanzini,
\textit{``{Wild Quiver Gauge Theories}''},
\doiref{10.1007/JHEP02(2012)031}{JHEP \textbf{1202}, 031
  (2012)\ignorespaces}\ignorespaces,
\normalsize{\texttt{\arxivref{1112.1691}{arXiv:1112.1691}}}\ignorespaces
\bibitem{Witten:2007td}
E.~Witten,
\textit{``{Gauge theory and wild ramification}''},
\normalsize{\texttt{\arxivref{0710.0631}{arXiv:0710.0631}}}\ignorespaces
\bibitem{Argyres:1995jj}
P.~C. Argyres \& M.~R. Douglas,
\textit{``{New phenomena in SU(3) supersymmetric gauge theory}''},
\doiref{10.1016/0550-3213(95)00281-V}{Nucl.~Phys. \textbf{B448}, 93
  (1995)\ignorespaces}\ignorespaces,
\normalsize{\texttt{\arxivref{hep-th/9505062}{hep-th/9505062}}}\ignorespaces
\bibitem{Argyres:1995xn}
P.~C. Argyres, M.~R. Plesser, N.~Seiberg \& E.~Witten,
\textit{``{New N=2 superconformal field theories in four-dimensions}''},
\doiref{10.1016/0550-3213(95)00671-0}{Nucl.~Phys. \textbf{B461}, 71
  (1996)\ignorespaces}\ignorespaces,
\normalsize{\texttt{\arxivref{hep-th/9511154}{hep-th/9511154}}}\ignorespaces
\bibitem{Wang:2018gvb}
Y.~Wang \& D.~Xie,
\textit{``{Codimension-two defects and Argyres-Douglas theories from
  outer-automorphism twist in 6d $(2,0)$ theories}''},
\normalsize{\texttt{\arxivref{1805.08839}{arXiv:1805.08839}}}\ignorespaces
\bibitem{Buican:2014hfa}
M.~Buican, S.~Giacomelli, T.~Nishinaka \& C.~Papageorgakis,
\textit{``{Argyres-Douglas Theories and S-Duality}''},
\doiref{10.1007/JHEP02(2015)185}{JHEP \textbf{1502}, 185
  (2015)\ignorespaces}\ignorespaces,
\normalsize{\texttt{\arxivref{1411.6026}{arXiv:1411.6026}}}\ignorespaces
\bibitem{Xie:2017vaf}
D.~Xie \& S.-T. Yau,
\textit{``{Argyres-Douglas matter and N=2 dualities}''},
\normalsize{\texttt{\arxivref{1701.01123}{arXiv:1701.01123}}}\ignorespaces
\bibitem{Wang:2015mra}
Y.~Wang \& D.~Xie,
\textit{``{Classification of Argyres-Douglas theories from M5 branes}''},
\doiref{10.1103/PhysRevD.94.065012}{Phys.~Rev. \textbf{D94}, 065012
  (2016)\ignorespaces}\ignorespaces,
\normalsize{\texttt{\arxivref{1509.00847}{arXiv:1509.00847}}}\ignorespaces
\bibitem{Xie:2017aqx}
D.~Xie \& K.~Ye,
\textit{``{Argyres-Douglas matter and S-duality: Part II}''},
\doiref{10.1007/JHEP03(2018)186}{JHEP \textbf{1803}, 186
  (2018)\ignorespaces}\ignorespaces,
\normalsize{\texttt{\arxivref{1711.06684}{arXiv:1711.06684}}}\ignorespaces
\bibitem{Beem:2013sza}
C.~Beem, M.~Lemos, P.~Liendo, W.~Peelaers, L.~Rastelli \& B.~C. van~Rees,
\textit{``{Infinite Chiral Symmetry in Four Dimensions}''},
\doiref{10.1007/s00220-014-2272-x}{Commun.~Math.~Phys. \textbf{336}, 1359
  (2015)\ignorespaces}\ignorespaces,
\normalsize{\texttt{\arxivref{1312.5344}{arXiv:1312.5344}}}\ignorespaces
\bibitem{Cecotti:2010fi}
S.~Cecotti, A.~Neitzke \& C.~Vafa,
\textit{``{R-Twisting and 4d/2d Correspondences}''},
\normalsize{\texttt{\arxivref{1006.3435}{arXiv:1006.3435}}}\ignorespaces
\bibitem{Argyres:2007cn}
P.~C. Argyres \& N.~Seiberg,
\textit{``{S-duality in N=2 supersymmetric gauge theories}''},
\doiref{10.1088/1126-6708/2007/12/088}{JHEP \textbf{0712}, 088
  (2007)\ignorespaces}\ignorespaces,
\normalsize{\texttt{\arxivref{0711.0054}{arXiv:0711.0054}}}\ignorespaces
\bibitem{Buican:2017rya}
M.~Buican \& Z.~Laczko,
\textit{``{Nonunitary Lagrangians and unitary non-Lagrangian conformal field
  theories}''},
\doiref{10.1103/PhysRevLett.120.081601}{Phys.~Rev.~Lett. \textbf{120}, 081601
  (2018)\ignorespaces}\ignorespaces,
\normalsize{\texttt{\arxivref{1711.09949}{arXiv:1711.09949}}}\ignorespaces
\bibitem{Gadde:2011uv}
A.~Gadde, L.~Rastelli, S.~S. Razamat \& W.~Yan,
\textit{``{Gauge Theories and Macdonald Polynomials}''},
\doiref{10.1007/s00220-012-1607-8}{Commun.~Math.~Phys. \textbf{319}, 147
  (2013)\ignorespaces}\ignorespaces,
\normalsize{\texttt{\arxivref{1110.3740}{arXiv:1110.3740}}}\ignorespaces
\bibitem{Rastelli:2016tbz}
L.~Rastelli \& S.~S. Razamat,
\textit{``{The supersymmetric index in four dimensions}''},
\doiref{10.1088/1751-8121/aa76a6}{J.~Phys. \textbf{A50}, 443013
  (2017)\ignorespaces}\ignorespaces,
\normalsize{\texttt{\arxivref{1608.02965}{arXiv:1608.02965}}}\ignorespaces
\bibitem{Gaiotto:2009we}
D.~Gaiotto,
\textit{``{N=2 dualities}''},
\doiref{10.1007/JHEP08(2012)034}{JHEP \textbf{1208}, 034
  (2012)\ignorespaces}\ignorespaces,
\normalsize{\texttt{\arxivref{0904.2715}{arXiv:0904.2715}}}\ignorespaces
\bibitem{Lemos:2014lua}
M.~Lemos \& W.~Peelaers,
\textit{``{Chiral Algebras for Trinion Theories}''},
\doiref{10.1007/JHEP02(2015)113}{JHEP \textbf{1502}, 113
  (2015)\ignorespaces}\ignorespaces,
\normalsize{\texttt{\arxivref{1411.3252}{arXiv:1411.3252}}}\ignorespaces
\bibitem{Witten:1982fp}
E.~Witten,
\textit{``{An SU(2) Anomaly}''},
\doiref{10.1016/0370-2693(82)90728-6}{Phys.~Lett. \textbf{B117}, 324
  (1982)\ignorespaces}\ignorespaces
\bibitem{DiPietro:2014bca}
L.~Di~Pietro \& Z.~Komargodski,
\textit{``{Cardy formulae for SUSY theories in $d =$ 4 and $d =$ 6}''},
\doiref{10.1007/JHEP12(2014)031}{JHEP \textbf{1412}, 031
  (2014)\ignorespaces}\ignorespaces,
\normalsize{\texttt{\arxivref{1407.6061}{arXiv:1407.6061}}}\ignorespaces
\bibitem{Buican:2015ina}
M.~Buican \& T.~Nishinaka,
\textit{``{On the superconformal index of Argyres-Douglas theories}''},
\doiref{10.1088/1751-8113/49/1/015401}{J.~Phys. \textbf{A49}, 015401
  (2016)\ignorespaces}\ignorespaces,
\normalsize{\texttt{\arxivref{1505.05884}{arXiv:1505.05884}}}\ignorespaces
\bibitem{Buican:2015hsa}
M.~Buican \& T.~Nishinaka,
\textit{``{Argyres–Douglas theories, S$^1$ reductions, and topological
  symmetries}''},
\doiref{10.1088/1751-8113/49/4/045401}{J.~Phys. \textbf{A49}, 045401
  (2016)\ignorespaces}\ignorespaces,
\normalsize{\texttt{\arxivref{1505.06205}{arXiv:1505.06205}}}\ignorespaces
\bibitem{Ardehali:2015bla}
A.~Arabi~Ardehali,
\textit{``{High-temperature asymptotics of supersymmetric partition
  functions}''},
\doiref{10.1007/JHEP07(2016)025}{JHEP \textbf{1607}, 025
  (2016)\ignorespaces}\ignorespaces,
\normalsize{\texttt{\arxivref{1512.03376}{arXiv:1512.03376}}}\ignorespaces
\bibitem{Buican:2017uka}
M.~Buican \& T.~Nishinaka,
\textit{``{On Irregular Singularity Wave Functions and Superconformal
  Indices}''},
\doiref{10.1007/JHEP09(2017)066}{JHEP \textbf{1709}, 066
  (2017)\ignorespaces}\ignorespaces,
\normalsize{\texttt{\arxivref{1705.07173}{arXiv:1705.07173}}}\ignorespaces
\bibitem{Cremonesi:2014xha}
S.~Cremonesi, G.~Ferlito, A.~Hanany \& N.~Mekareeya,
\textit{``{Coulomb Branch and The Moduli Space of Instantons}''},
\doiref{10.1007/JHEP12(2014)103}{JHEP \textbf{1412}, 103
  (2014)\ignorespaces}\ignorespaces,
\normalsize{\texttt{\arxivref{1408.6835}{arXiv:1408.6835}}}\ignorespaces
\bibitem{Kapustin:2010xq}
A.~Kapustin, B.~Willett \& I.~Yaakov,
\textit{``{Nonperturbative Tests of Three-Dimensional Dualities}''},
\doiref{10.1007/JHEP10(2010)013}{JHEP \textbf{1010}, 013
  (2010)\ignorespaces}\ignorespaces,
\normalsize{\texttt{\arxivref{1003.5694}{arXiv:1003.5694}}}\ignorespaces
\bibitem{Aharony:2013dha}
O.~Aharony, S.~S. Razamat, N.~Seiberg \& B.~Willett,
\textit{``{3d dualities from 4d dualities}''},
\doiref{10.1007/JHEP07(2013)149}{JHEP \textbf{1307}, 149
  (2013)\ignorespaces}\ignorespaces,
\normalsize{\texttt{\arxivref{1305.3924}{arXiv:1305.3924}}}\ignorespaces
\bibitem{Gaiotto:2010jf}
D.~Gaiotto, N.~Seiberg \& Y.~Tachikawa,
\textit{``{Comments on scaling limits of 4d N=2 theories}''},
\doiref{10.1007/JHEP01(2011)078}{JHEP \textbf{1101}, 078
  (2011)\ignorespaces}\ignorespaces,
\normalsize{\texttt{\arxivref{1011.4568}{arXiv:1011.4568}}}\ignorespaces
\bibitem{Tachikawa:2018rgw}
Y.~Tachikawa, Y.~Wang \& G.~Zafrir,
\textit{``{Comments on the twisted punctures of $A_\text{even}$ class S
  theory}''},
\normalsize{\texttt{\arxivref{1804.09143}{arXiv:1804.09143}}}\ignorespaces
\bibitem{Aharony:2013hda}
O.~Aharony, N.~Seiberg \& Y.~Tachikawa,
\textit{``{Reading between the lines of four-dimensional gauge theories}''},
\doiref{10.1007/JHEP08(2013)115}{JHEP \textbf{1308}, 115
  (2013)\ignorespaces}\ignorespaces,
\normalsize{\texttt{\arxivref{1305.0318}{arXiv:1305.0318}}}\ignorespaces
\bibitem{Argyres:2016yzz}
P.~C. Argyres \& M.~Martone,
\textit{``{4d $ \mathcal{N} $ =2 theories with disconnected gauge groups}''},
\doiref{10.1007/JHEP03(2017)145}{JHEP \textbf{1703}, 145
  (2017)\ignorespaces}\ignorespaces,
\normalsize{\texttt{\arxivref{1611.08602}{arXiv:1611.08602}}}\ignorespaces
\bibitem{Inaki}
I.~Garc\'ia-Etxebarria,
\textit{``New $\CN=4$ Theories in Four Dimensions (Talk at Strings 2017)''}
\bibitem{Lin:2004nb}
H.~Lin, O.~Lunin \& J.~M. Maldacena,
\textit{``{Bubbling AdS space and 1/2 BPS geometries}''},
\doiref{10.1088/1126-6708/2004/10/025}{JHEP \textbf{0410}, 025
  (2004)\ignorespaces}\ignorespaces,
\normalsize{\texttt{\arxivref{hep-th/0409174}{hep-th/0409174}}}\ignorespaces
\end{thebibliography}

\end{document}